\documentstyle[preprint,aps,epsfig]{revtex}
	
\newcommand {\be} {\begin{equation}} 
\newcommand {\ee} {\end{equation}} 
\newcommand {\Be}{\begin{eqnarray*}}
\newcommand {\Ee} {\end{eqnarray*}}
\newcommand {\bey} {\begin{eqnarray}} 
\newcommand {\eey} {\end{eqnarray}}

\begin{document}
\begin{center}

\Large{\bf Heat conduction in 2d nonlinear lattices}\\

\vspace{0.5cm}

{\large Andrea Lippi}\\

\vspace{0.2cm}
{\large and}

\vspace{0.2cm}
{\large Roberto Livi$^{\dagger}$ }\\
{\small\it
Dipartimento di Fisica dell'Universit\`a\\
L.go E.Fermi 2 I-50125 Firenze, Italy\\
and Istituto Nazionale di Fisica della Materia, Unit\`a di Firenze \\
\noindent
{\rm lippi@fi.infn.it}, {\rm livi@fi.infn.it}\\
}

\end{center}

\date{\today}
\begin{abstract}
\noindent
{
The divergence of the heat conductivity in the thermodynamic limit is
investigated in 2d-lattice models of anharmonic solids with 
nearest-neighbour interaction from single-well potentials. Two different 
numerical approaches 
based on nonequilibrium and equilibrium simulations provide consistent 
indications in favour of a logarithmic divergence in "ergodic", i.e. highly 
chaotic, dynamical regimes.
Analytical estimates obtained in the framework of 
linear-response theory confirm this finding, while tracing back 
the physical origin of this {\sl anomalous} transport to the slow 
diffusion of the energy of long-wavelength effective Fourier modes. 
Finally, numerical evidence of {\sl superanomalous} transport 
is given in the weakly chaotic regime, typically found below 
some energy density threshold.
}

\vspace{0.5cm}

\vspace{0.3cm}
\end{abstract}
\vfill
$^{\dagger}$ Istituto Nazionale di Fisica Nucleare, Sezione di Firenze 

\newpage
\section{Introduction}

The study of heat conduction in models of insulating solids 
is a long-standing and debated problem. In 1929 R. Peierls 
provided the first convincing theoretical explanation of this
phenomenon, relying upon the hypothesis that lattice vibrations
responsible for heat transport in insulating solids can be described
as a diluted gas of quantum quasi-particles, the {\sl phonons}
\cite{peierls}. Peierls' model extends Boltzmann's kinetic
theory of the classical ideal gas to the phonon gas, by
substituting collisions in real space with {Umklapp} processes
in momentum space. Due to the quantum character of this approach,
Umklapp processes were theoretically justified as a perturbative
effect, epitomizing the scattering of phonons originated by
intrinsic features of real solids: disorder and nonlinearity.
The main achievement of Peierls' theory is the prediction of
the proportionality relation between the thermal conductivity
$\kappa$ and the specific heat at constant volume $c_V$:
\begin{equation}
\kappa(T)  = \frac{1}{3} c_V(T) v \lambda
\label{kcv} 
\end{equation}
where $v$ and $\lambda$ are the average velocity and the mean free path
of phonons, respectively. Using Debye's formula, the dependence
of $\kappa$ on the temperature $T$ was found to agree with
experimental observations in the low-temperature regime.
Nonetheless, at least in the high temperature regime, there is no 
reason why a purely classical model of an insulating solid should not
reproduce the correct behaviour of $\kappa(T)$ and Fourier's law
\begin{equation}
{\vec J}(\vec x) = \kappa  \vec\nabla T(\vec x)
\label{Fourier} 
\end{equation}
where $\vec J (\vec x)$ and $T(\vec x)$ are the heat flux and the 
temperature field, respectively.
To our knowledge, the first 
attempt of its theoretical justification on the basis of a microscopic
model of a solid was made in a seminal paper by Rieder, Lebowitz 
and Lieb \cite{lebo}. They considered a chain of $N$ coupled 
harmonic oscillators 
in contact at its extrema with stochastic thermal baths at different 
temperatures. They prove that in this simple model of a purely
harmonic solid thermal conductivity diverges in the thermodynamoc
limit as $\kappa (N) \sim N$.
Consistently,  the temperature profile in
the bulk of the chain is flat. The physical interpretation
of these results is that harmonic lattice waves,
i.e. Fourier's modes, freely propagate through the lattice,
thus contributing to a ballistic rather than diffusive transmission
of heat.
The same result was obtained numerically for another integrable
model, the 1d Toda lattice \cite{mokro}. Its normal modes, the 
so-called {\sl Toda solitons}, are nonlinear waves localized in space
and also freely propagating through the lattice.
More generally, any integrable Hamiltonian system is expected to
be characterized by anomalous transport properties since its 
normal modes, in the perspective of kinetic theory, are equivalent
to a "gas" of non-interacting free particles.
Following Peierls, the addition of generic disorder and nonlinearity 
should overtake these anomalies by introducing effective diffusive 
behaviour. Numerical \cite{ishi} and analytical \cite{papa} 
estimates performed for the mass-disordered harmonic chain still yield 
anomalous transport, $\kappa(N) \sim N^{1/2}$, although the temperature 
gradient was found to be nonzero \cite{lebo1}. 
Accordingly, disorder
is not sufficient to guarantee normal transport properties.
It remains nonlinearity, that since the 70's has been widely studied
also as a source of chaotic behaviour. In fact, despite its deterministic
nature, chaos was shown to yield good statistical properties, namely 
ergodic behaviour, for most of the thermodinamically relevant
observables. Accordingly, many attempts were made with models of chains 
of nonlinearly coupled oscillators, that did not achieve convincing
conclusions (e.g., see\cite{naka,payt,viss}).
Finally, Casati et al. \cite{casa} proposed the ding-a-ling model
that exhibits normal heat conductivity. It is worth mentioning that 
the same result has been confirmed one decade later by Prosen and Robnik
\cite{pros} for the ding-dong model - a variant of the ding-a-ling one.
The great merit of the former paper has been to point out the possibility
of obtaining normal conductivity in a nonlinear dynamical model;
the latter, thanks also to the increase of computational
facilities, has provided convincing evidence of this
result by very careful and time consuming numerical simulations.
Anyway, in both cases strong chaos has been argued to be the 
basic mechanism responsible for normal transport properties. In this
perspective, one could conjecture that the doubtful results obtained 
for chaotic potentials, yielding interactions smoother than elastic
collisions could deserve extremely lengthy numerical simulations
before showing standard transport properties (e.g., see\cite{mist,kabu}).

Unfortunately, the explanation of the above mentioned
results is quite different. Both ding-models can be thought as
interacting particles bounded to a local, or {\sl substrate}, potential.
By substituting the elastic collisions - responsible for
the strong chaotic behaviour of these models - with any nonlinear
nearest-neighbour interaction potential, numerical evidence of finite 
thermal conductivity is achieved (for instance, see ref.\cite{bamb}). 
Conversely, if the
substrate term is absent small wavenumber, i.e. {\sl hydrodynamic}, 
modes, as a consequence of the conservation of momentum , 
exhibit a peculiar slow relaxation \cite{alab}. This effect can
be traced back to the vanishing of the generalized frequency
$\omega (k)$ in the limit of small wavenumber, $k\to 0$, characterizing the
dispersion relation in presence of an acoustic band \cite{LiviEu}.
In practice, the frequency associated with the local potential lifts the 
acoustic band from zero, thus making possible the diffusion of the
{\sl hydrodynamic} modes energy.
This notwithstanding, it is certainly
non-trivial that a purely deterministic chaotic dynamics can give
rise to an effective diffusive behaviour. Nonetheless, chaos alone is 
a necessary but not sufficient ingredient for obtaining
normal trasport properties
\footnote{
We want to point out that normal thermal conductivity has
been recently observed also in a class of 1d hamiltonian models
without any substrate potential and intercating via nearest-neighbour
periodic and double-well chaotic potentials \cite{giar}.}. 

Some progress in the understanding of transport properties in
nonlinear lattices has been made recently by exploiting the analogy 
with the mode-coupling theory, that applies to models of dense fluids
rather than diluted gases. Numerical simulations 
carried out for a 1d lattice with quartic nonlinearity -
the Fermi-Pasta-Ulam (FPU) $\beta$-model - have shown a power law
dependence of the thermal conductivity on the system size: 
$\kappa \sim N^{\alpha}$, with $\alpha = 0.39 \pm 0.02$ \cite{LiviEu}. 
This result was obtained for a chain
in contact with thermal baths and by equilibrium measurements
of $\kappa$ based on the Green-Kubo formula of linear response
theory (for technical details , see also Sec.4). Taking advantage of 
the results reported in \cite{alab}, it was possible to conclude that,
even for strongly chaotic dynamics, the amplitudes of low wavenumber
modes evolve like 
weakly damped stochastic harmonic oscillators, whose frequencies are
{\sl renormalized} w.r.t. the standard harmonic component of the
FPU $\beta$-model by leading nonlinear terms \cite{Livi1}. 
On the other hand,
large wavenumber modes behave like {\sl fast variables}, governed by a 
{\sl thermal} behaviour. This is the typical scheme that applies 
to dense fluids in Fourier space. Self-consistent mode coupling theory
(SMT) \cite{pomeau,ernst}
allows one  to determine the dependence of $\kappa$ on $N$ from the integral
of the time-correlation function of the total heat flux, yielding in the
1d case the power law
\begin{equation}
\kappa \sim N^{2/5}
\label{scal}
\end{equation}
in very good agreement with numerics.  This power-law has been 
recovered also in many other models of nonlinearly coupled oscillators
like the diatomic Toda chain \cite{hata,vass}, the Lennard-Jones (LJ) 6/12 
model and the Morse potential \cite{lipp}. All of these models share
one common feature: they are single-well confining potentials.
Also in the light of the results recently obtained in \cite{giar},
the scaling law (\ref{scal}), typical of
this class of models, seems to be directly related to this property.
It is also worth stressing that, due to its generality, SMT can be 
extended to determine the dependence of transport coefficients on the
system size in higher space dimensions.
For the class of models of interest in this paper, it
predicts that $\kappa$ should diverge logarithmically with $N$ in 2d
(see Section 4), while in 
3d $\kappa$ is normal, i.e. independent of $N$.

The aim of this paper is the numerical and analytical study 
of the thermodynamic limit properties of thermal conductivity
$\kappa$ in the 2d case, by taking into account two examples of
single-well confining potentials:
the FPU $\beta$ and the LJ-(6/12) models. 

More precisely, in Section 2 we introduce the models and we also
discuss some general features of the simplified lattice representation
of anharmonic 2d solids used in this paper. In Section 3 we present
the results concerning the dependence of $\kappa$ on $N$ obtained
by two different numerical approaches, based on two different
definitions of $\kappa$. An explicit derivation of the analytical
prediction of SMT is worked out in Section 4, where we also
discuss the presence of a {\sl superanomalous} divergence
of transport coefficients in the weakly chaotic regime, that
typically characterizes the above mentioned models for sufficiently small
energy \cite{bene} (see also \cite{case,giar1}). Conclusions and
perspectives are contained in Section 5.
\section{ Modelling heat conduction in 2d lattices}

Most of the studies on the
heat conduction problem in anharmonic lattices has been devoted
to 1d systems. Time-consuming numerical simulations were the main
limitation for large scale analysis of this problem in 2d.
In fact, only a few contributions have been worked out.
For instance in \cite{moun} the dependence of the thermal conductivity
$\kappa$ on the temperature $T$ is studied in a 2d triangular lattice of 
unit mass atoms interacting via a LJ 6/12 potential.
Numerical data are consistent with the expected classical law 
$\kappa \sim T^{-1}$, while the dependence 
of $\kappa$ on the system size is not investigated.
An interesting contribution in this direction is the paper by
Jackson and Mistriotis \cite{mist}. The authors compare measurements
of $\kappa$ in 1d and 2d Fermi-Pasta-Ulam (FPU) lattices and 
they conclude that in both cases there is no evidence that the transport 
coefficient is finite in the thermodynamic limit.
More recent and extended numerical simulations performed 
for the 2d Toda-lattice \cite{NKS} have been interpreted in
favour of the finiteness of $\kappa$ in this case.

As usual, numerical analysis alone without any piece of a theory can
hardly yield conclusive results. On the other hand, as already 
stressed in \cite{mist},
the dependence of $\kappa$ on the system size cannot be adequately described 
by Peierls' model: 
at least in the high temperature (i.e., classical) limit.
the perturbative Umklapp processes 
cannot account for the genuine nonlinear effects that characterize 
such a dependence. 

The theoretical approach proposed in \cite{LiviEu} and 
careful numerical analysis of 1d systems \cite{Livi1,Livi2},
provide a first coherent explanation of the divergence 
of $\kappa(N)$ in the thermodynamic limit. 

In this paper we want to extend this study to 2d systems.
For this purpose, let us first describe
the adopted lattice model of a homogeneous 2d solid. 
We consider a square lattice made
of $N_x \times N_y$ equal mass ($m$) atoms. 
The equilibrium positions of the atoms coincide with the lattice sites, 
labelled by a pair of integer indices $(i,j)$. 
Without loss of generality, the lattice spacing $a$ can
be set equal to unit and the origin of the cartesian reference frame 
can be fixed in
such a way that $1< i < N_x$ and $1< j < N_y$.
Accordingly, the 2d-vector of equilibrium position, ${\vec r}^0_{ij}$, 
coincides with $(i,j)$.
The {\sl short-range} character of interatomic forces in real
solids is simplified by assuming that the atoms interact by
a nearest-neighbour confining potential $V(\rho)$, that depends on 
the relative displacement $\rho$ with respect to the equilibrium distance.
The natural ordering of the atoms induced by this kind of local interaction 
allows to identify with the same couple of indices $(i,j)$ the corresponding 
atom in any dynamical configuration.
Specifically, the model is described by the general Hamiltonian
\begin{equation}
H = \sum_{i=1}^{N_x} \sum_{j=1}^{N_y} \left[ \left(\frac{|{\vec p}_{ij}|^2}
{2m}\right) + V(|\vec q_{i+1j} - \vec q_{ij}|) + V(|\vec q_{ij+1} - \vec q_{ij}|)
\right]
\label{Ham}
\end{equation}
where $\vec q_{ij}(t) = \vec r_{ij}(t) - \vec r^0_{ij}$,
$\vec r_{ij}(t)$ is the instantaneous position vector of the $(i,j)$-atom
and $\vec p_{ij}(t)$ is
the corresponding momentum vector. 
The time dependence of the canonical variables has been 
omitted in eq.(\ref{Ham}). 
Without prejudice of generality we set also $m = 1$.

Since we are interested in studying the thermal conductivity $\kappa$ 
in this 2d model, we have also to define the relevant physical observables,
namely the temperature $T_{ij}$ and the heat flux $\vec J_{ij}$.
Let us start from the local energy density
\begin{equation}
h(\vec r,t) = \sum_{i,j} h_{ij} \delta(\vec r -\vec r_{ij})
\label{dens}
\end{equation}
where
\begin{eqnarray}
h_{ij} &=& \frac{1}{2}\left|\vec p_{ij}\right|^2 + \frac{1}{4} [V(|\vec q_{i+1j}- 
\vec q_{ij}|) + V(|\vec q_{ij} - \vec q_{i-1j}|)\nonumber\\ 
&+& V(|\vec q_{ij+1} - \vec q_{ij}|) + V(|\vec q_{ij} - \vec q_{ij-1}|)]
\label{dh} 
\end{eqnarray}
Assuming that local equilibrium holds we obtain immediately a definition
of the local temperature $T_{ij}$ by applying the virial theorem:
\begin{equation}
T_{ij} = \langle p^{(k)}_{ij}\frac{\partial h_{ij}}{\partial p^{(k)}_{ij}}\rangle = 
\langle q^{(k)}_{ij}\frac{\partial h_{ij}}{\partial q^{(k)}_{ij}}\rangle 
\label{vir}
\end{equation}
where $k = x,y$ indicates either the $x$- or the $y$-component of
the corresponding vector variables. The formal definition
of the time average symbol is
$\langle \bullet \rangle = \lim_{\tau\to\infty} \frac{1}{\tau}
\int_0^{\tau} \bullet \quad dt$. In numerical experiments the time-average
must be performed over a finite time span , in general much larger that 
the local equilibration time. 

For hamiltonian (\ref{Ham}) one can use the simple expression
\begin{equation}
T_{ij} = \langle {p^{(x)}_{ij}}^2 \rangle = \langle {p^{(y)}_{ij}}^2 \rangle = 
{ \langle {{p^{(x)}_{ij}}^2 + {p^{(y)}_{i,j}}^2}\rangle \over 2}
\label{temp}
\end{equation}
The heat flux vector is implicitely defined by the continuity equation
\begin{equation}
\dot h_{i,j}(t) + \vec\nabla \cdot \vec J_{ij}(t) = 0
\label{cont}
\end{equation}
By rewriting this equation in Fourier-space and retaining the leading
hydrodynamic contribution (i.e., by applying the large wavelenght limit)
one obtains an explicit expression for the components of $\vec J$
\begin{eqnarray}
J^{(x)}_{ij} &= - \frac {1}{4} a\left[f^{xx}_{ij}\left(p^{(x)}_{ij} + 
p^{(x)}_{i+1j}\right) + f^{yx}_{ij}\left(p^{(y)}_{ij} + p^{(y)}_{i+1j}\right)\right] \cr
J^{(y)}_{ij} &= - \frac {1}{4} a\left[f^{xy}_{ij}\left(p^{(x)}_{ij} + 
p^{(x)}_{ij+1}\right) + f^{yy}_{ij}\left(p^{(y)}_{ij} + p^{(y)}_{ij+1}\right)\right]
\end{eqnarray}
where the components of the local vector forces are defined as follows
$$ 
f^{xx}_{ij} = - \frac{\partial V \left[\left| \vec q_{i+1j} - \vec q_{ij}\right|\right]}
{\partial q^{(x)}_{ij}} \quad f^{yx}_{ij} = - \frac{\partial V \left[\left| \vec q_{i+1j} 
- \vec q_{ij}\right|\right]}{\partial q^{(y)}_{ij}}$$
$$ 
f^{xy}_{ij} = - \frac{\partial V \left[\left| \vec q_{ij+1} - \vec q_{ij}\right|\right]}
{\partial q^{(x)}_{ij}} \quad f^{yy}_{ij} = - \frac{\partial V \left[\left| \vec q_{ij+1} 
- \vec q_{ij}\right|\right]}{\partial q^{(y)}_{ij}}$$

It is worth defining also the space-time average of the heat flux vector
\begin{equation}
\langle {\vec J} \rangle = \langle \frac{\sum_{i,j} {\vec J}_{ij}}{N_x N_y}
\rangle
\label{avflux}
\end{equation}
that will be widely used in the following Sections.

Finally, since we are interested to investigate thermodynamic limit 
properties, we should impose periodic boundary conditions ({\sl pbc}),
that are the standard choice for reducing as much as possible
undesired boundary effects on bulk properties.

\section{ Models and numerical experiments}

The numerical simulations that we are going to describe in this
Section have been performed for two different single-well potentials:

\begin{equation}
V_1(\rho) = \frac{1}{2} \rho^2 + \frac{\beta}{4} \rho^4 
\quad\quad ,
\quad\quad Fermi-Pasta-Ulam \quad \beta-model 
\label{FPUpot}
\end{equation}
\begin{equation}
V_2(\rho) = \frac{A}{\rho^{12}} - \frac{B}{\rho^6} + \frac {B^2}{4A}
\quad\quad ,
\quad\quad Lennard-Jones \quad 6/12 \quad model
\label{LJpot}
\end{equation}

The reasons for this twofold choice are the following:
\begin{itemize}

\item[-]analyzing heat tansport in the 2d version of the widely
studied case of potential $V_1$;

\item[-]verifying the theoretical prediction (see next Section)
that nearest-neighbour single-well nonlinear potentials exhibit the same
kind of dependence of $\kappa(N)$ in the large $N$ limit;

\item[-]investigating the possible consequences on transport 
properties of the slowing-down of relaxation below the so called
{\sl Strong Stochasticity Threshold}, that was measured for 
potential $V_2$ in \cite{bene}.
\end{itemize}

It is worth stressing that $V_1$ does not contain any
natural energy and length scales; all numerical simulations 
with this potential have been performed with $\beta = 10^{-1}$ 
and with an integration time step $\Delta t = 10^{-2}$, that
is two orders of magnitude smaller than the minimum harmonic
period ($\tau_{min} = \pi/\sqrt{2}$).

At variance with $V_1$, $V_2$ is characterized by natural 
length and energy scales, the equilibrium distance
$r_0 = (2A/B)^{1/6}$ and the well depth $W = B^2/4A$, respectively. 
In order to make the two models as close
as possible we have chosen the parameters $A$ and $B$ in such a way
that the coefficients of the second and fourth order terms of the Taylor 
series expansion of $V_2$ around its minimum coincide with those of 
$V_1$, thus obtaining $r_0 \approx 25$, $W \approx 8.6$ and $\tau_{min}
= \pi r_0^4/(6\sqrt{2} B)
\approx 2.2$. In  numerical experiments we have adopted a time 
step $\Delta t = 5\cdot 10^{-3}$, that guarantees a sufficient sampling
of the integration algorithm especially when strong nonlinearities
are explored by the dynamics.

Numerical measurements of the thermal conductivity $\kappa$ as a function
of the lattice size $N$ have been performed under nonequilibrium and 
equilibrium conditions: the  former are inspired by an ideal experiment 
for the verification of Fourier's law (\ref{Fourier}), the latter stem 
from linear response theory (see Sec. 4).
Both approaches are theoretically equivalent; nonetheless,
their comparison on a numerical ground is quite
instructive. 

\begin{figure}
\centering\epsfig{figure=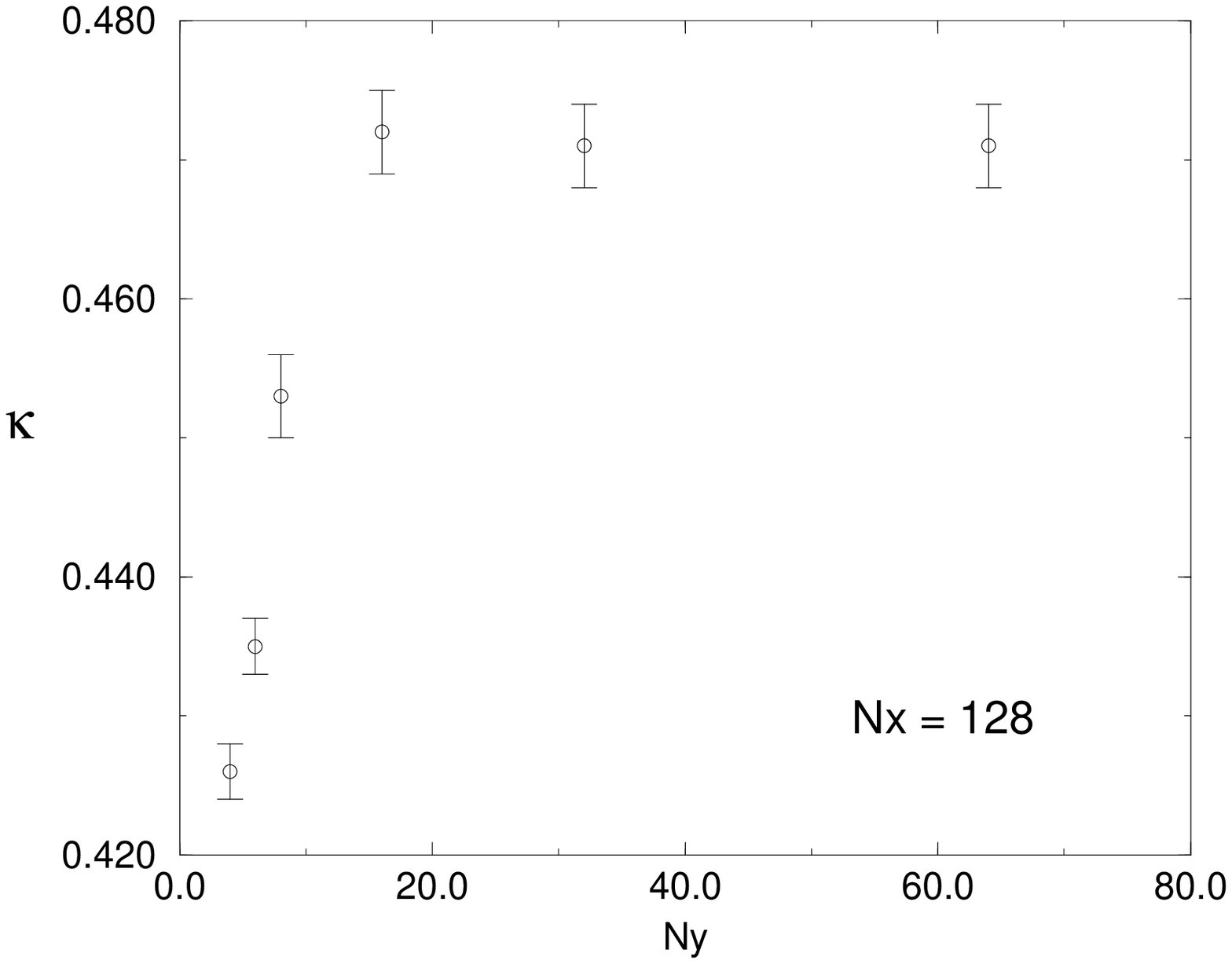,width=11. cm}
\caption{The thermal conductivity $\kappa$ versus $N_y$ for the
Lennard-Jones potential (\ref{LJpot}) with $N_x = 128$,
$T_L = 1.$ and $T_R = 0.5$~.}
\label{fluxtras}
\end{figure}

Nonetheless, this kind of numerical simulations
are quite heavy, so that any trick for saving CPU time is
worth to be applied. 
Since we are interested in investigating thermodynamic
limit properties we have to perform measurements
keeping the ratio $R = N_y/N_x$ constant. On the other
hand, there is {\sl a priori} no reason for choosing
$R=1$. In fact, we have checked that
reliable measurements of $\kappa$ can be obtained in lattices 
with $ R < 1$. 
As an example, in Fig.\ref{fluxtras} we show the dependence
of $\kappa$ on $N_y$ for $N_x = 128$ and for potential $V_2$,
in the presence of thermal baths acting on the boundary atoms
at temperatures $T_L = 1.$ and $T_R = 0.5$ (details are given in the
next subsection).  After a sharp
increase for small values of $N_y$ thermal conductivity $\kappa$ 
reaches a plateau already for $N_y \approx 20$.

This preliminary analysis has been performed for both potentials
$V_1$ and $V_2$ in nonequilibrium and in equilibrium simulations: 
we have concluded that
$R = 1/2$ is a reasonable compromise, valid in all of these cases.

\subsection{Nonequilibrium measurements of thermal conductivity}

A straightforward way of measuring thermal conductivity 
$\kappa$ amounts
to simulate numerically a true experiment where the atoms at the
left and right edges of the 2d lattice are coupled with two thermal
baths at different temperatures $T_L$ and $T_R$.
This setting
restricts the application of {\sl pbc} to the direction orthogonal to
the temperature gradient:
we impose {\sl pbc} along the y-axis, while the boundary atoms
are coupled to a rigid wall through the force link with the
missing atom in the $x$-direction.

Various models of thermal baths, either stochastic or deterministic
ones, are at disposal. The results of numerical simulations reported
in this paper have been obtained by using the Nos\'e-Hoover deterministic 
model \cite{Nose,Hoover} \footnote{
We have checked that other models of thermal baths, 
in particular stochastic ones, yield the 
same kind of results reported in this paper.}. 
It has two advantages with respect to stochastic algorithms:
it can be easily implemented as an ordinary differential
equation and it reduces the residual thermal 
impedence effects at the lattice boundaries. 

The equations of motion are
\begin{eqnarray}
\dot {\vec q}_{ij} &=& {\vec p}_{ij} 
\cr
\dot {\vec p}_{ij} &=& -{\partial V\over{\partial {\vec q}_{ij}}} 
- (\delta_{i,1} + \delta_{i,N_x}) \zeta_{ij} {\vec p}_{ij} 
\cr
\dot\zeta_{1j} &=& \frac{1}{\theta^2}\left[{|{\vec p}_{1j}|^2 
\over {2T_L}} -1\right]
\cr
\dot\zeta_{N_xj} &=& \frac{1}{\theta^2}\left[{|{\vec p}_{N_xj}|^2 
\over {2T_R}} -1\right]
\label{algo}
\end{eqnarray}
where $\delta$ is the Kronecker symbol. In practice, the integration 
scheme (\ref{algo}) has been implemented by a standard fourth order 
Runge-Kutta algorithm.
Note that each boundary atom is coupled through the momentum equation
to its thermal bath variable $\zeta$, that guarantees
local thermal equilibrium at temparature $T_L$ and $T_R$ on the
left and right boundaries, respectively.

Starting from random initial conditions after a sufficiently
long transient time stationary nonequilibrium evolution is eventually
approached.  According to the heat equation (\ref{Fourier}), a  
constant thermal gradient should establish through the lattice in the
x-direction and $\langle J^{(x)} \rangle > 0 $, while $\langle
J^{(y)} \rangle = 0$. Here, the symbol $\langle \bullet \rangle$ indicates
the time average over stationary nonequilibrium states.
The time span, over which good convergence of the time-average is obtained,
increases with $N_x$: for instance, ${\cal O}(10^5)$ integration steps are 
sufficient for $N_x = 16$, while for $N_x = 128$ this time grows up to
${\cal O}(10^7)$. 

The occurrence of stationary nonequilibrium conditions can be checked 
by verifying the equality
\begin{equation}
|\langle \vec J \rangle| = 
\langle J_M \rangle =  - \langle \frac{1}{N_y}\sum_{j=1}^{N_y}
\zeta_{Mj} |{\vec p}_{Mj}|^2 \rangle \quad\quad , \quad\quad M = 1,N_x
\label{avbflux}
\end{equation}
where the r.h.s. is the average heat flux flowing through the boundaries.
Despite in numerical simulations finite time-averages never yield exactly
$\langle J^{(y)} \rangle = 0$, one finds that  
$|\langle \vec J \rangle|$ is very well approximated by
$\langle \vec J^{(x)} \rangle$.

\begin{figure}
\centering\epsfig{figure=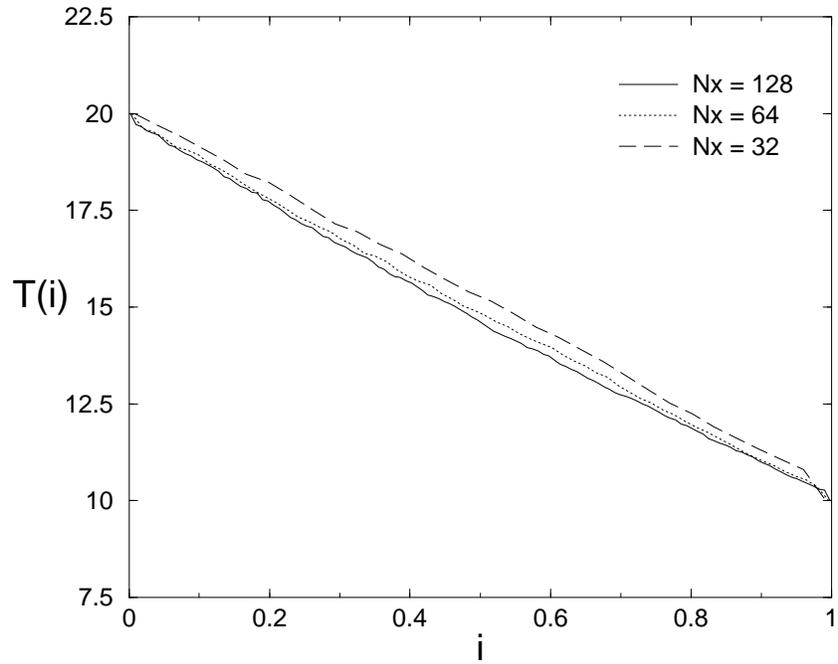,width=11. cm}
\caption{Temperature profiles of the 2d FPU-$\beta$ 
model for different values of $N_x$}
\label{proFPU}
\end{figure}

\begin{figure}
\centering\epsfig{figure=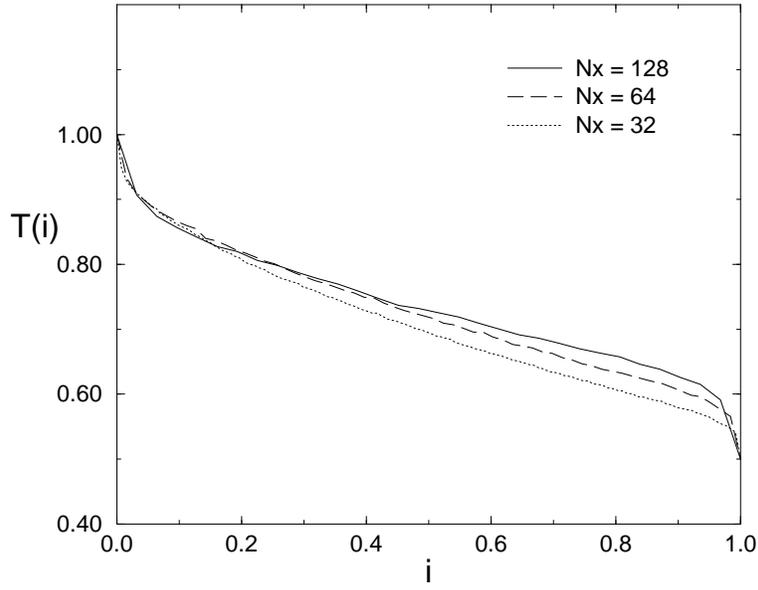,width=11. cm}
\caption{Temperature profiles of the 2d Lennard-Jones
6/12 model model for different values of $N_x$}
\label{proLJ}
\end{figure}

Some of the stationary temperature profiles obtained for different values 
of $N_x$ for potential $V_1$ and $V_2$ are shown in Figures \ref{proFPU} and
\ref{proLJ}, respectively. The values of the temperature $T_i$
have been averaged also in space over all the $N_y$ atoms with abscissa
$x = i$: 
\begin{equation}
T_i = \frac{1}{N_y} \sum_{j=1}^{N_y} T_{ij}
\end{equation}
It is worth stressing that the local temperature $T_{ij}$ is a
time-averaged quantity (see eq.(\ref{temp})) over the stationary 
nonequilibrium evolution.

The lattice length 
has been rescaled to unit in order to verify the overlap of
the profiles for increasing values of $N_x$: in both cases we observe
a good data collapse for $N_x > 32$ indicating that the temperature
gradient in the thermodynamic limit vanishes as
\begin{equation}
\vec\nabla T_i \sim N_x^{-1}
\label{gradsc}
\end{equation}

\begin{figure}
\centering\epsfig{figure=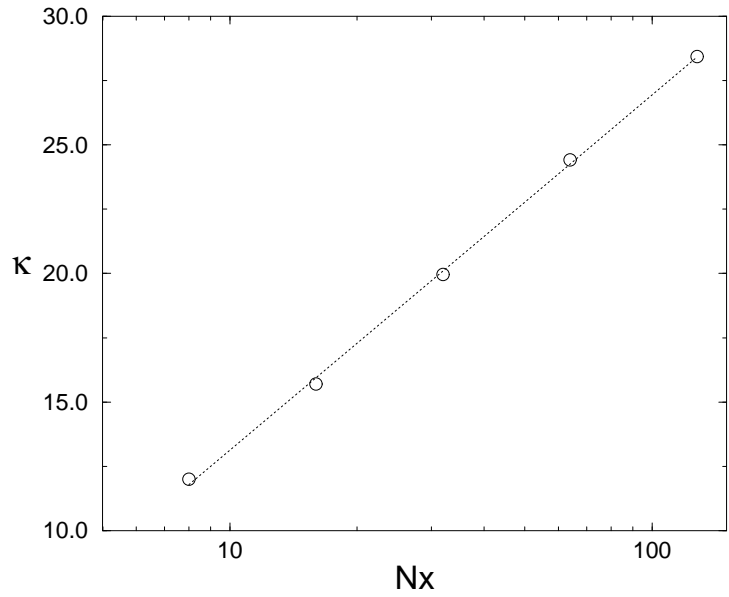,width=11.cm}
\caption{Dependence of the thermal conductivity 
$\kappa$ on the system size $N_x$ for the FPU $\beta$-model; 
$T_L =20.$ and $T_R = 10.$~. 
Statistical errors have the size of the symbols and the dotted line
is the best fit.}
\label{kappafpu}
\end{figure}

\begin{figure}
\centering\epsfig{figure=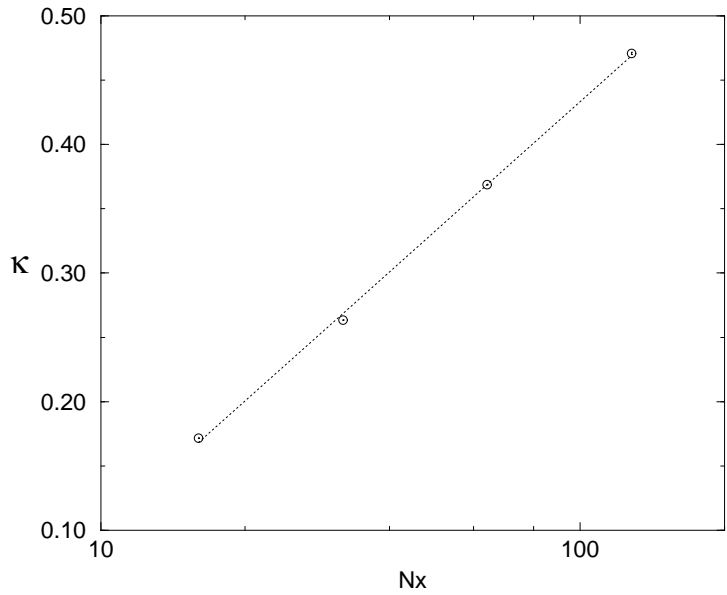,width=11.cm}
\caption{Dependence of the thermal conductivity $\kappa$
on the system size $N_x$ for the Lennard-Jones (6/12)-model;
$T_L = 1.$ and $T_R = 0.5$.
Statistical errors have the size of the symbols and the dotted line
is the best fit.}
\label{kappalj}
\end{figure}

In the FPU $\beta$-model the temperatures of the thermal baths,
$T_L = 20$ and $T_R = 10$,
have been chosen for obtaining good ergodic properties of the dynamics,
i.e. fast relaxation to local equilibrium. Boundary effects 
of thermal impedence induced by the coupling with the thermal baths
are reduced for increasing values of $N_x$.
The temperature gradient looks quite close to
a constant, but a more careful inspection shows
that the profile has a slight curvature.  
The effect is even more evident for the Lennard-Jones
model, whose temperature profiles converge to an s-shaped curve.
In particular, thermal impedence at the boundaries
seems to persist also in the large $N_x$ limit, despite the smaller
values of $T_L = 1$ and $T_R = 0.5$, that already guarantee 
ergodicity for this model. 
Such deviations from Fourier's law 
indicate that nonlinearities influence boundary effects and also
the dependence of $\kappa$ on the temperature. 
On the other hand, we are interested in extracting the scaling
of $\kappa$ with the system size $N_x$. 
It can be obtained
on the basis of Fourier's law and eq.(\ref{gradsc}) by plotting 
the quantity $\langle J^{(x)} \rangle N_x \sim \kappa $ versus $N_x$
(see Figures  \ref{kappafpu} and \ref{kappalj}).
For both 2d models we find evidence of a logarithmic scaling of
$\kappa$ with $N_x$.  

\subsection{Equilibrium measurements of thermal conductivity}

The above described nonequilibrium measurements of the
heat transport coefficient $\kappa$ have
been made keeping $T_L$ and $T_R$ constant for increasing
values of $N_x$. The data collapse of the temperature profiles implies
also that in the thermodinamic limit, $N_x \to \infty$, the 
temperature gradient in the bulk of the chain drops to zero.
Accordingly, for increasing values of $N_x$ nonequilibrium measurements 
are expected to reproduce better and better linear response conditions.
The Green-Kubo linear response theory \cite{kubo} provides an alternative
definition of $\kappa$ with respect to the one contained in
Fourier's law. More precisely, a general expression of the 
heat conductivity tensor for a solid contained in a volume
$V$ is given by the formula
\begin{equation}
\kappa_{\mu\nu} = \frac{V}{K_B T^2} \int_0^{\infty}
\langle J^{(\mu)}(t) J^{(\nu)}(0) \rangle dt
\label{GK}
\end{equation}
where $K_B$ is the Boltzmann constant
and the time correlation function of the heat flux components
along the $\mu$ and $\nu$ directions is averaged over equilibrium
states at temperature $T$. Note that the symbol $\langle 
\bullet \rangle$ indicates now the equilibrium ensemble average.

In our models of homogeneous 2d nonlinear solids the thermal
conductivity corresponds indifferently to anyone of the diagonal
components of the transport coefficient tensor defined in
(\ref{GK}). In particular we have chosen to measure

\begin{equation}
\kappa \equiv \kappa_{xx} = \frac{R N_x^2}{K_B T^2} 
\lim_{t\to\infty}\int_0^t
\langle J^{(x)}(\tau) J^{(x)}(0) \rangle d\tau
\label{GKx}
\end{equation}
where $J^{(x)}$ is the $x$-component of the space-averaged
heat flux vector (\ref{avflux}). As in nonequilibrium measurements 
fixing the ratio $R = N_y/N_x = 1/2$ is a good compromise
for reducing CPU time, while maintaining reliable asymptotic
estimates of $\kappa$.

We have performed numerical simulations at constant energy
density, $e$, by eliminating the thermal 
bath variables $\zeta$'s from eq.(\ref{algo}).
Periodic boundary conditions have been imposed also in the 
$x$-direction, so that the 2d lattice becomes a torus.
In practice the integration scheme has been implemented by 
a fourth order Maclachlan-Atela
symplectic algorithm \cite{macl}, that is more appropriate than a
Runge-Kutta one for this kind of {\sl microcanonical}
simulations.

\begin{figure}
\centering\epsfig{figure=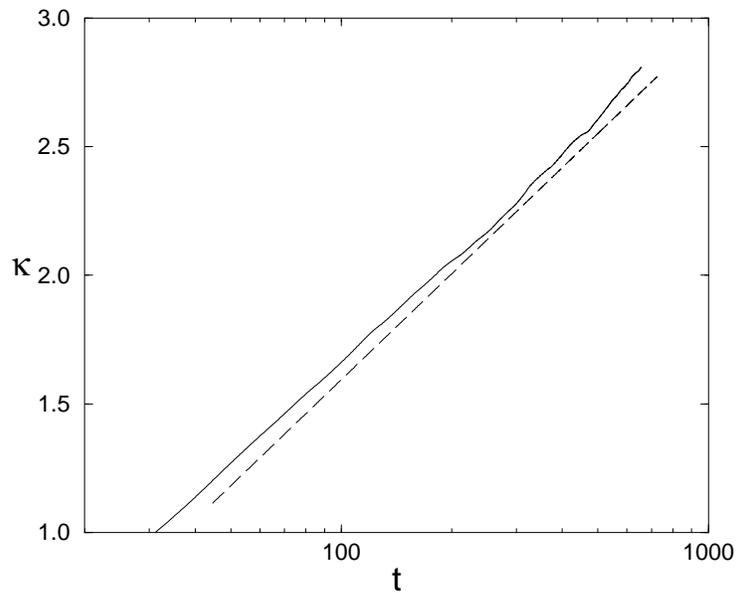,width=11.cm}
\caption{The finite-time thermal conductivity $\kappa(t)$ versus time $t$ 
for the LJ (6/12)-model with $N_x = 64$ and $e = 1.5$~.
The curve has been obtained by averaging over 100 initial conditons.
The dashed straight line has been drawn just for comparison with numerical
data.
}
\label{Ljkubo}
\end{figure}

The expression on the r.h.s of eq.(\ref{GKx}) can be evaluated
numerically for finite values of $t$. Accordingly, we define
a "finite time" thermal conductivity $\kappa (t)$ by removing from
eq.(\ref{GK}) the limit operation $t \to \infty$.
In order to reduce fluctuations, the heat flux 
time-correlation function has been averaged over
a sufficiently large set of random initial conditions,
extracted from the microcanonical probability distribution.
For sufficiently large values of $t$ both models (\ref{FPUpot}) and
(\ref{LJpot}) exhibit a logarithmic divergence of $\kappa (t)$.
As an example, in Figure \ref{Ljkubo} we report the case of the 
Lennard-Jones (6-12)-model. 
Despite highly time consuming simulations have been performed for averaging 
over initial conditions, in this example
fluctuations are still persistent for large values of $t$.
This notwithstanding, numerical data are compatible with the expected 
logarithmic divergence.

According to the argument reported in \cite{LiviEu} this divergence
in time can be assimilated to a divergence with the system size
$N_x$. In fact, numerical estimates of the time-correlation
function of the local heat flux, $C_i(\tau) = \langle 
J_{0j}(\tau) J_{ij}(0) \rangle $
show that in single-well nonlinear potentials, due to the
"rigidity" of low-wavenumber modes (see next Section), the energy
propagates with the velocity of sound in a nonlinear medium
(see also \cite{alab}):
\begin{equation}
\tilde c = \sqrt{(1 + \alpha (E))} c
\label{vel}
\end{equation}
where $c$ is the velocity of sound due to the harmonic
component of the nonlinear potential and the renormalization
factor $\alpha(E)> 0$ introduces a dependence of $\tilde c$ on
the total energy, indicating its nonlinear character.
We have verified that these features are common to the
potentials (\ref{FPUpot} ) and (\ref{LJpot} ).

The evaluation  of
$\kappa(t)$ for finite time $t$ in the Green-Kubo integral can
be assimilated to an estimate of the divergence with the system size
by the formal relation $t = N_x/{\tilde c}$. In this sense, equilibrium
simulations confirm the results of nonequilibrium ones.

\section{Transport in strong and weak chaotic dynamics}
The conjecture that deterministic chaos is an efficient mechanism
for ergodic behaviour has been extensively investigated 
by numerical experiments
in many degrees of freedom Hamiltonian systems,
like those introduced in Section 2. One of the main issues is
that, for sufficiently high energy density $e$, the 
time averages of most observables of thermodynamic interest 
rapidly approach the expectation value predicted by 
equilibrium ensembles. This notwithstanding, below a specific
value $e_T$ - the so-called Strong 
Stochasticity Threshold ({\sl SST}) (see \cite{case,giar1} and the references
therein contained) - the equilibration time may increase
dramatically, despite the persistence of deterministic chaos.
It is not our aim, here, to discuss why such a slow relaxation
mechanism typically occurs in these models.
We just want to stress that their {\sl mild } nonlinear character 
make them quite different from {\sl mathematically} standard chaotic
models like K- or A-systems.
In fact, the results obtained in \cite{alab} show that regularities 
are present even above the SST. More precisely, in a 1d FPU chain 
the amplitudes of low-$k$ Fourier modes are found to evolve like weakly 
damped and forced harmonic oscillators. Dissipation and forcing epitomize
the complex nonlinear mechanism of energy exchange among the modes
and they are found to vanish in the limit $k \to 0$.
One can argue that these can be only an effect induced by global,
i.e. {\sl hydrodynamic}, conservation laws of total momentum and energy.
In this sense, such a behaviour is expected to be present in any space
dimension $d$, although the possibilities of energy exchanges among the
modes are expected to become more and more efficient for increasing
values of $d$.
These remarks suggests that, in a
nonlinear model of a solid, Fourier modes can be assimilated to 
a dense fluid rather than to a diluted gas, as in Peierls' 
phonon theory. 

\subsection{Analytical estimate of anomalous thermal conductivity above the SST}
Relying upon the above conjecture, strongly supported by numerical
simulations, we want to derive an analytical estimate of the
thermodynamic limit behaviour of thermal conductivity $\kappa$
for 2d single-well anharmonic lattices. In practice, here we
explicitely extend to the 2d case the method introduced in \cite{LiviEu} 
and described in detail for the 1d case in \cite{lepri}.

The amplitudes of the Fourier modes in a 2d square lattice have the
standard  expression
\begin{equation}
{\vec Q}_{\vec k} = \frac {1}{\sqrt{N_xN_y}} \sum_{m=1}^{N_x} 
\sum_{n=1}^{N_y} {\vec q}_{mn} e^{-i\left(\frac{2\pi}{N_x}k_xm 
+ \frac{2\pi}{N_y}k_yn\right)}
\label{Mod}
\end{equation}
where ${\vec q}_{mn}$ is the canonical space coordinate and the
2d wave-vector ${\vec k}$ has components  
$k_x = -\frac{N_x}{2} \dots \frac{N_x}{2}$ and 
$k_y = -\frac{N_y}{2} \dots \frac{N_y}{2}$.
We consider an isolated hamiltonian model of the type (\ref{Ham});
the equations of motion expressed in terms of Fourier amplitudes
and their canonically conjugated momenta ${\vec P}_{\vec k}$ read
\begin{eqnarray}
\dot{\vec Q}_{\vec k} &=& {\vec P}_{\vec k} \cr
\dot{\vec P}_{\vec k} &=& -\omega^2_{\vec k} {\vec Q}_{\vec k} 
+ \sum_{\vec k \not= {\vec k}^{\prime}} {\vec F}_{\vec k {\vec k}^{\prime}}
\label{Ppunto}
\end{eqnarray}
where $\omega_{\vec k}$ obeys the dispersion relation
\footnote{For he sake of simplicity we have set 
$\omega_0 = \left(\frac{\partial^2 V}{\partial r^2}\right)_{r=r_0} = 1$}
\begin{equation}
\omega^2_{\vec k} = 4 \left[ \sin^2 \frac{\pi k_x}{N_x} + \sin^2 \frac{\pi k_y}{N_y}\right]
\label{Disper}  
\end{equation} 
$F_{\vec k {\vec k}^{\prime}}$ is the formal expression of the nonlinear
interaction force between modes $\vec k$ and ${\vec k}^{\prime}$.

Numerical experiments show that low-$k$ Fourier modes behave like
{\sl slow} dynamical variable, whose relaxation time scales
are much longer than those of high-$k$ modes, that play the role
of {\sl fast} variables. This analogy with hydrodynamic properties
of dense fluids suggests the application of SMT
\cite{pomeau,ernst}.

Since the hydrodynamic behaviour is ruled by low-$k$ modes,
the first step amounts to project the dynamics onto the 
subspace of slow variables 
$({\vec Q}^{s}_{\vec k}, {\vec P}^{s}_{\vec k})$ 
by a proper projection operator
\begin{equation}
{\cal P}^{s}X = \sum_{\vec k}\left[ \frac{\langle X \cdot
{\vec Q}^{s*}_{\vec k}\rangle}
{\langle\left|{\vec Q}^{s}_{\vec k} \right|^2\rangle} 
{\vec Q}^{s}_{\vec k} 
+ \frac{\langle X{\vec P}^{s*}_{\vec k}\rangle}{\langle\left| 
{\vec P}^{s}_{\vec k} 
\right|^2\rangle}{\vec P}^{s}_{\vec k}\right]
\label{Mori}
\end{equation}
According to linear response theory the equations of motion can be casted in
the form \cite{lepri}: 
\begin{eqnarray}
\dot {\vec Q}^{s}_{\vec k} &=& {\vec P}^{s}_{\vec k} \cr
\dot {\vec P}^{s}_{\vec k} &=& -\tilde\omega^2_{\vec k} 
{\vec Q}^{s}_{\vec k} - \int_0^t \Gamma_{\vec k}(t-t') 
{\vec P}^{s}_{\vec k}(t')dt' + {\vec R}_{\vec k}
\label{projec}
\end{eqnarray}
where $ {\vec R}_{\vec k} = (1 - {\cal P}^{s}) \dot {\vec P}^{s}_{\vec k} $
is the effective force, that is related by the dissipation-fluctuation
theorem to the memory kernel as follows:
\begin{equation}
\Gamma_{\vec k}(t) \propto \langle {\vec R}_{\vec k}(t) \cdot
{\vec R}_{\vec k}(0)
\rangle
\label{Flucdis}
\end{equation}
Note that in the projected equations of motion the harmonic frequency
$\omega_{\vec k}$ is renormalized to the energy dependent 
frequency
\begin{equation}
\tilde\omega_{\vec k} = \sqrt{(1+ \alpha(E))}\omega_{\vec k}
\label{Renor}
\end{equation}
where
\begin{equation}
\alpha(E) = \frac{1}{\beta} \frac{\int \delta(H-E) d \vec r}{\int r^2 
\delta(H-E) d \vec r} -1
\label{alfa}
\end{equation}
In the last formula we have introduced the microcanonical average
by assuming that equilibrium properties are practically guaranteed
by the fast thermalization of high-$k$ modes.
As observed also in \cite{alab} the projected slow variables 
can be interpreted as "nonlinear" Fourier modes, whose frequency
is renormalized by anharmonic contributions. Numerical simulations
support this approach also thanks to the
remarkable quantitative agreement with the theoretical 
prediction (\ref{Renor}).

The projected equations of motion have a clear theoretical interpretation
but they are practically useless for working out analytical calculations.
We have to introduce some simplifying hypotheses. 
We suppose that the time scale of
slow hydrodynamic variables can be unambigously separated form the typical
microscopic time scales: specifically, this amounts to reduce the
memory kernel $\Gamma$ to a $\delta$ distribution and the fluctuating
force $\vec R$ to a stochastic force.
By introducing an explicit complex-variable representation of the
mode amplitudes, 
$${\vec Q}_{\vec k} = {\vec A}_{\vec k} + i {\vec B}_{\vec k}$$
we rewrite the set of equations (\ref{projec}) in the approximate 
form of standard stochastic equations:
\begin{eqnarray}
\ddot {{\vec A}_{\vec k}} + \gamma_{\vec k} \dot {{\vec A}_{\vec k}} 
+ \tilde\omega^2_{\vec k} {\vec A}_{\vec k} &= {\vec \xi}_{\vec k} \cr
\ddot {{\vec B}_{\vec k}} + \gamma_{\vec k} \dot {\vec B_{\vec k}} + 
\tilde\omega^2_{\vec k} {\vec B}_{\vec k} &= {\vec \eta}_{\vec k}
\label{ABkappa}
\end{eqnarray}
where $\gamma_{\vec k}$ can be interpreted as an effective dissipation
coefficient, while fluctuations are now assumed to be represented by
gaussian white noise processes: 
\begin{eqnarray}
\langle{\vec \xi}_{\vec k}(t) {\vec \xi}_{\vec k'}(t') \rangle 
&\propto \delta_{{\vec k}{\vec k'}} \delta (t - t') \cr
\langle{\vec \eta}_{\vec k}(t) {\vec \eta}_{\vec k'}(t')\rangle 
&\propto \delta_{{\vec k}{\vec k'}} \delta (t - t')
\end{eqnarray}
It is convenient to introduce the scalar variable
\begin{equation}
W_{\vec k} = \dot{\vec A}_{\vec k} \cdot \vec B_{\vec k} - 
\vec A_{\vec k} \cdot \dot{\vec B}_{\vec k}
\label{vudoppio}
\end{equation}
that satisfies the Langevin type equation:
\begin{equation}
\dot W_{\vec k} = -\gamma_{\vec k} W_{\vec k} + \zeta_{\vec k}
\label{Lang}
\end{equation}
where 
$$ \langle \zeta_{\vec k}(t) \zeta_{\vec k'}(t') \rangle 
\propto (|\vec A_{\vec k}|^2 + |\vec B_{\vec k}|^2 ) 
\delta_{\vec k \vec k'} \delta (t-t')$$ 
According to the definition (\ref{avflux}) the average heat flux
vector can be thought as the sum of a harmonic term and an
anharmonic one, ${\vec J_H}$ and ${\vec J_A}$, respectively.
The harmonic term is obtained by considering only the 
forces given by the quadratic part of the interaction
potential: simple calculations show that its components
can be written in the form:
\begin{eqnarray}
J_H^{(x)} &= \sum_{k_x,k_y} c_{k_x} \omega_{k_x} W_{\vec k} \cr
J_H^{(y)} &= \sum_{k_x,k_y} c_{k_y} \omega_{k_y} W_{\vec k}
\end{eqnarray}
where
$$ \omega_{k_{x,y}} = 2\left| \sin \frac{\pi k_{x,y}}{N_{x,y}} \right|
\quad\quad , \quad\quad 
c_{k_{x,y}} =\frac{N_{x,y}}{\pi} \frac{d \omega_{k_{x,y}}}{d k_{x,y}}$$ 
In close analogy with the renormalization of the effective frequency
of slow variables, it can be shown that the leading contribution 
to ${\vec J_A}$, stemming from the anharmonic part of the interaction
potential, is proportional to ${\vec J_H}$ through an energy (and also model)
dependent factor $C(E)$, i.e. ${\vec J_A} = C(E) {\vec J_H}$.
For instance, the expression of $C(E)$ in the FPU $\beta$-model
(\ref{FPUpot}) is
\begin{equation}
\frac{1}{N_xN_y} \sum_{\vec k} \omega^2_{\vec k}\left( \left| 
\vec A_{\vec k} \right|^2 + \left| \vec B_{\vec k} \right|^2 \right)
\end{equation}

Summarizing, the components of the average heat flux 
vector can be expressed in general by the proportionality 
relations
\begin{eqnarray}
J^{(x)} &\propto \sum_{k_x,k_y} c_{k_x} \omega_{k_x} W_{\vec k} \cr
J^{(y)} &\propto \sum_{k_x,k_y} c_{k_y} \omega_{k_y} W_{\vec k}
\end{eqnarray}
Assuming the validity of the generalized equipartiton theorem
\begin{equation}
\omega^2_{\vec k} \langle |A_{\vec k}|^2 \rangle = \omega^2_{\vec k} 
\langle |B_{\vec k}|^2 \rangle = {\cal U}(E)
\label{equip}
\end{equation}
where ${\cal U}(E)$ is a function of the energy $E$, and using
the solution of (\ref{Lang}), one can obtain an analytical
estimate of the heat flux time-correlation function 
present in the Green-Kubo integral (\ref{GK}):
\begin{equation}
\langle J^i_H (t) J^i_H(0) \rangle 
\sim {\cal U}(E) \sum_{\vec k} 
c^2_{k_i} e^{-\gamma_{\vec k} t} \sim \int_0^{2\pi} d\theta \int_0^{\pi} 
\frac{dk}{(2\pi)^2} k c^2(k \cos \theta) e^{-\gamma (k)t}
\label{Corr}
\end{equation}
where the last expression is obatined by assuming that in the thermodynamic
limit the summation can be approximated by an integral. 

The explicit dependence of the dissipation coefficient $\gamma$ 
on the modulus of the wave-vector $\vec k$ 
is provided by SMT \cite{pomeau,ernst}.
Specifically, it predicts that the time-correlation functions
of hydrodynamic modes decay in time as follows:
\begin{equation}
G(k,t)  \simeq e^{- \omega(k) t}
\end{equation}
where the $\omega (k)$ are complex generalized frequencies
ruling the oscillatory and relaxation behaviour of the hydrodynamic modes.
For 2d homogeneous systems they depend just on 
$k = |\vec k|$ and their explicit expression in the hydrodynamic
limit, $k \to 0$, is found to be
\begin{equation}
\omega(k) \simeq i \tilde c k + \nu k^2 \ln k
\end{equation}
where $\tilde c$ is the velocity of sound in the homogeneous medium
and the second addendum of the r.h.s. is the dissipation term
$\gamma (k) = {\cal R}e(\omega(k))$.  
Substituting into eq.(\ref{Corr}) and retaining only the leading
contribution to the function
$c(k \cos \theta)$ in the limit $k \to 0$, one obtains the
estimate
\begin{equation}
\langle J^i_H (t) J^i_H(0) \rangle \propto
\int_0^{\pi} \frac{dk}{2\pi} k e^{-tk^2 \ln k} \sim \frac {1}{t} 
+ {\cal O}\left( \frac{1}{t \ln t} \right)
\label{correc}
\end{equation}
Accordingly, the heat conductivity $\kappa$ is predicted to
diverge as $\ln t$ in the $t \to \infty$ limit in 2d lattices.

It is worth stressing that SMT predicts also that
$\gamma(k) \sim k^{5/3}$ in 1d systems, yielding a 
diverging heat conductivity $\kappa \sim N^{2/5}$ in the
thermodynamic limit. The very good agreement with numerical 
experiments (see \cite{LiviEu}) strengthen the conjecture that
the anomalous properties of transport coefficients in nonlinear 
lattice models stems from the hydrodynamic nature of low-$k$
effective modes. Said differently, total energy and momentum
conservation laws impose on models with single-well potentials a
dispersion relation yielding a subdiffusive behaviour of the 
energy exchange among these modes and, consequently, ill-defined
transport coefficients. On the other hand, SMT predicts
that in 3d the Green-Kubo integral is convergent so that $\kappa$
becomes a well defined quantity.

This dependence on the space dimension and the generality of 
predictions (at least for what concerns the class of 
single-well nonlinear potentials) make this piece of theory quite 
elegant and physically sound.

\subsection{The effect of weak chaos on anomalus transport}
All the numerical simulations presented in Section 3 have been performed in the
strong chaotic regime of the dynamics. In this case 
ergodic behaviour occurs quite rapidly on the microscopic time
scale, with the exception of hydrodynamic modes, whose slow relaxation,
due to macroscopic conservation laws, rules the
observed divergence of transport coefficients.
In the above theoretical treatment 
this corresponds to the assumption that time-correlation functions
can be estimated on the basis of equilibrium ensemble averages.

On the other hand, we know that below the SST relaxation to
equilibrium may slow-down dramatically for most of the
physically interesting observables. According to the results
reported in \cite{giar1} the heat flux is expected to belong 
to this class of observables. Then, we expect also that in such a 
dynamical regime the transport mechanism can be significantly modified.

\begin{figure}
\centering\epsfig{figure=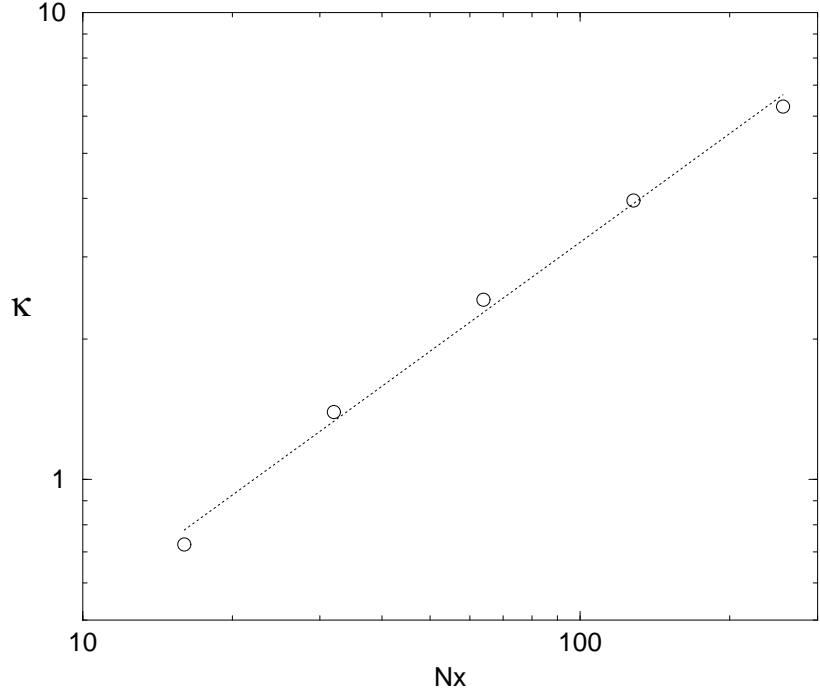,width=11. cm}
\caption{The thermal conductivity $\kappa$ versus the system size $N_x$ with 
$T_L = 0.1$ and $T_R = 0.05$ for the LJ (6/12)-model. The dotted line is a 
best fit for the power law $\kappa \sim N_x^{\alpha}$ yielding 
$\alpha \simeq 0.77 \pm 0.03$~.}
\label{kappasst}
\end{figure}

We have checked this conjecture by considering model (\ref{LJpot}),
whose SST has been estimated \cite{bene} at a value of the energy 
density
\begin{equation}
e_{SST} = 0.3
\end{equation}
in the units adopted in this paper.
We have performed numerical simulations with thermal baths
at temperature $T_L = 0.1$ and $T_R = 0.05$ for different
values of $N_x$ verifying the data collapse of the
temperature profiles. Following the same approach described
in Section 3.1 we have estimated the dependence of the thermal
conductivity on $N_x$. We find evidence of a power-law divergence
\begin{equation}
\kappa \sim  N_x^{\alpha}
\end{equation}
with $\alpha =0.77$ (see Fig.\ref{kappasst}). Note that below the SST the 
exponent $\alpha$ is a function of the energy density $e$.

\begin{figure}
\centering\epsfig{figure=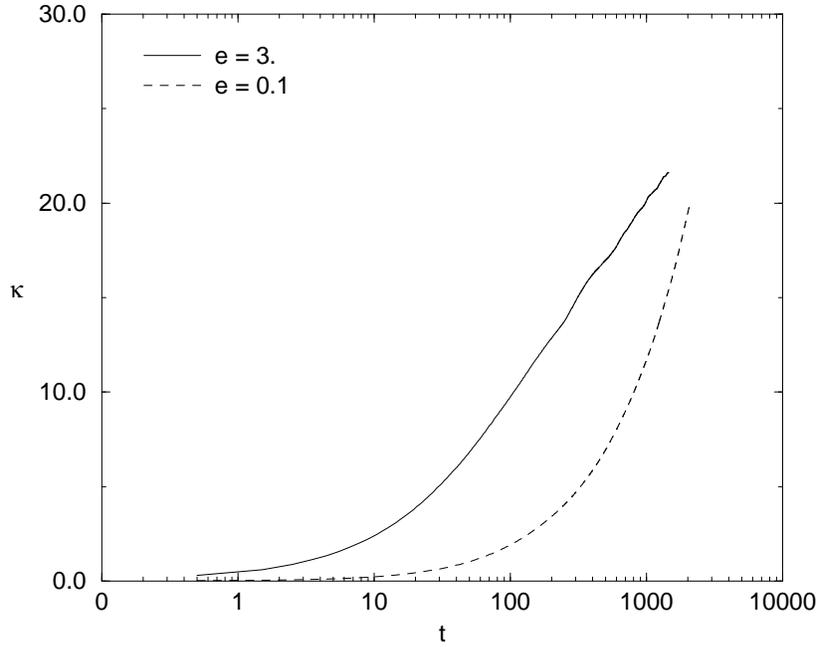,width=11.cm}
\caption{The finite-time thermal conductivity $\kappa$ versus time $t$ 
for the LJ (6/12)-model with $N_x = 64$ below ($e = 0.1$)
and above ($e = 3.0$) the SST.
}
\label{gkubosst}
\end{figure}

The power law divergence of $\kappa$ below the SST is confirmed 
by the numerical estimate of the Green-Kubo integral.
In Fig.~\ref{gkubosst} we compare dependence of $\kappa$ on
time $t$ for $e = 0.1$
and for $e = 3.0$~, below and above 
the SST, respectively. The comparison of the two curves in a
linear versus logarithmic scale shows that 
the asymptotic behaviour of $\kappa(t)$ is completely different
in the two chaotic dynamical regimes. In particular, 
for $e = 3.0$ we recover a logarithmic divergence, while
for $e = 0.1$ 
we find a power law divergence $\kappa(t) \sim t^{\alpha}$
with $\alpha = 0.75$. 

For what concerns the interpretation of this last result, one
has to observe that for a sufficiently long time equilibrium
conditions will be eventually approached. The asymptotic
time dependence of the Green-Kubo integral should turn to
the logarithmic growth on any finite system. 
The relevance of this {\sl superanomalous} effect could be
established only by evaluating the dependence of the relaxation time
on the system size. Estimates obtained for relatively small size
systems seem to indicate that the crossover time between the power-law
and the logarithmic behaviour of the Green-Kubo integral
icreases more than linearly with the system size $N_x$.
On the other hand, as we have often pointed out along this paper,
numerics may can provide crucial insight for the understanding of 
these phenomena,
but cannot be assumed as a proof of anything.
Since this effect might have very interesting physical consequences
we hope to work out in the near future an analytical approach
apt to describe the {\sl superanomaly} of transport coefficients
below the SST. 
\section{Conclusions and perspectives}

The presence of anomalous thermal conductivity in 2d lattices
of atoms coupled by nonlinear nearest-neighbour single-well 
potentials has been verified by numerical experiments and 
analytical estimates. According to SMT, the effect of dimensionality
is found to make the thermodynamic limit power-law divergence of 1d models
turn to a logarithmic divergence of
2d models, thus confirming the soundness of the general theoretical 
framework worked out to interpret such anomalous properties 
emerging from strongly chaotic dynamics \cite{LiviEu,lepri}.
It is also worth mentioning that in this dynamical regime 3d models 
are predicted to exhibit normal transport properties, i.e. finite
thermal conductivity. 

In this paper we have also pointed out that a superanomalous behaviour
ruled by a power-law divergence of the thermal conductivity, seems to 
characterize transport properties of 2d models below the SST.
Such a behaviour has been observed also in 1d systems where 
the crossing of the SST corresponds to an increase of the power
from the value 2/5 towards the limit value 1, that is expected
to be approached for vanishing energy densities (i.e. in the
harmonic limit). Here we have decided to report just the results
for 2d systems, where the comparison between the power-law and
the logarithmic divergence stresses the effect of
the extremely slow relaxation mechanism already observed almost
half a century ago in the seminal numerical experiment by
E. Fermi, J. Pasta and S. Ulam \cite{FPU}.

At present we are not able to conclude just on the basis of numerical
simulations if superanomalous transport is a finite
size effect or an asymptotic property, that could even concern
3d systems. Also in this case, the construction of a suitable theoretical
approach would greatly help in answering to this interesting question.

\vskip .3 cm
\centerline{\bf Acknowledgments}

\noindent
We want to thank S. Lepri, A. Politi and M. Vassalli for 
many useful discussions and suggestions.
One of the authors (R.L.) wants to acknowledge also the kind hospitality 
of the Institute for Scientific Interchange in Torino 
during the workshop "Complexity and Chaos" 1999, when 
part of this work has been performed.

\end{document}